\def\year{2021}\relax
\documentclass[letterpaper]{article} 
\usepackage{aaai}  
\usepackage{times}  
\usepackage{helvet} 
\usepackage{courier}  
\usepackage[hyphens]{url}  
\usepackage{graphicx} 
\usepackage{natbib}  
\usepackage{caption} 
\usepackage[super]{nth}
\usepackage{parskip}

\usepackage[utf8]{inputenc}
\usepackage[T1]{fontenc}
\usepackage{hyphenat}
\usepackage{xspace}
\usepackage{amsmath}
\usepackage{amsfonts}
\usepackage{url}
\usepackage{booktabs}
\usepackage{multirow}
\usepackage{subfigure}
\usepackage{makecell}
\usepackage{caption}
\usepackage{minibox}
\usepackage{bbm}
\usepackage{graphicx}
\usepackage{balance}
\usepackage{mathtools}
\usepackage{color}
\usepackage{marvosym}
\usepackage{ifthen}
\usepackage{textcomp}
\usepackage{enumitem}
\usepackage{verbatim}
\usepackage{algorithm}
\usepackage{algorithmic}
\usepackage{numprint}

\usepackage{amsthm}
\theoremstyle{plain}

\definecolor{MyRed}{rgb}{0.6,0.0,0.0} 
\definecolor{MyBlack}{rgb}{0.1,0.1,0.1} 
\newcommand{\inred}[1]{{\color{MyRed}\sf\textbf{\textsc{#1}}}}

\newcommand{\frameit}[2]{
  \begin{center}
  {\color{MyRed}
  \framebox[.9\columnwidth][l]{
    \begin{minipage}{.85\columnwidth}
    \inred{#1}: {\sf\color{MyBlack}#2}
    \end{minipage}
  }\\
  }
  \end{center}
}

\newcommand{\chatoDisplayMode}[1]{#1}

\newcommand{\note}[2][]{\chatoDisplayMode{\def\@tmpsig{#1}\frameit{{\Pointinghand} Note}{#2\ifx \@tmpsig \@empty \else \mbox{ --\em #1}\fi}}}
\newcommand{\todo}[2][]{\chatoDisplayMode{\def\@tmpsig{#1}\frameit{{\Writinghand} To-do}{#2\ifx \@tmpsig \@empty \else \mbox{ --\em #1}\fi}}}

\newcommand{\abbrevStyle}[1]{#1}

\newcommand{\ie}{\abbrevStyle{i.e.}\xspace}
\newcommand{\eg}{\abbrevStyle{e.g.}\xspace}
\newcommand{\cf}{\abbrevStyle{cf.}\xspace}

\newcommand{\vs}{\abbrevStyle{vs.}\xspace}
\newcommand{\etc}{\abbrevStyle{etc.}\xspace}

\newcommand{\Secref}[1]{Sec.~\ref{#1}}
\newcommand{\Eqnref}[1]{Eq.~\ref{#1}}

\newcommand{\Tabref}[1]{Table~\ref{#1}}
\newcommand{\Figref}[1]{Fig.~\ref{#1}}

\newcommand{\Appref}[1]{Appendix~\ref{#1}}

\newcommand{\xhdr}[1]{\vspace{1.7mm}\noindent{{\bf #1.}}}

\newcommand{\textcite}[1]{\citeauthor{#1} \shortcite{#1}}

\newcommand{\cpt}[1]{\textsc{\MakeLowercase{#1}}}

\newcommand{\hide}[1]{}

\makeatletter
\newcommand{\iffont}[2]{\ifthenelse{\equal{\f@family}{#1}}{#2}{}}
\makeatother

  \usepackage{mathptmx}

  \DeclareSymbolFont{greek}{OML}{cmm}{m}{n}
  \DeclareMathSymbol{\alpha}{\mathalpha}{greek}{"0B}
  \DeclareMathSymbol{\beta}{\mathalpha}{greek}{"0C}
  \DeclareMathSymbol{\gamma}{\mathalpha}{greek}{"0D}
  \DeclareMathSymbol{\delta}{\mathalpha}{greek}{"0E}
  \DeclareMathSymbol{\epsilon}{\mathalpha}{greek}{"0F}
  \DeclareMathSymbol{\zeta}{\mathalpha}{greek}{"10}
  \DeclareMathSymbol{\eta}{\mathalpha}{greek}{"11}
  \DeclareMathSymbol{\theta}{\mathalpha}{greek}{"12}
  \DeclareMathSymbol{\iota}{\mathalpha}{greek}{"13}
  \DeclareMathSymbol{\kappa}{\mathalpha}{greek}{"14}
  \DeclareMathSymbol{\lambda}{\mathalpha}{greek}{"15}
  \DeclareMathSymbol{\mu}{\mathalpha}{greek}{"16}
  \DeclareMathSymbol{\nu}{\mathalpha}{greek}{"17}
  \DeclareMathSymbol{\xi}{\mathalpha}{greek}{"18}
  \DeclareMathSymbol{\pi}{\mathalpha}{greek}{"19}
  \DeclareMathSymbol{\rho}{\mathalpha}{greek}{"1A}
  \DeclareMathSymbol{\sigma}{\mathalpha}{greek}{"1B}
  \DeclareMathSymbol{\tau}{\mathalpha}{greek}{"1C}
  \DeclareMathSymbol{\upsilon}{\mathalpha}{greek}{"1D}
  \DeclareMathSymbol{\phi}{\mathalpha}{greek}{"1E}
  \DeclareMathSymbol{\chi}{\mathalpha}{greek}{"1F}
  \DeclareMathSymbol{\psi}{\mathalpha}{greek}{"20}
  \DeclareMathSymbol{\omega}{\mathalpha}{greek}{"21}
  \DeclareMathSymbol{\varepsilon}{\mathalpha}{greek}{"22}
  \DeclareMathSymbol{\vartheta}{\mathalpha}{greek}{"23}
  \DeclareMathSymbol{\varpi}{\mathalpha}{greek}{"24}
  \DeclareMathSymbol{\varrho}{\mathalpha}{greek}{"25}
  \DeclareMathSymbol{\varsigma}{\mathalpha}{greek}{"26}
  \DeclareMathSymbol{\varphi}{\mathalpha}{greek}{"27}
  \DeclareSymbolFont{otone}{OT1}{cmr}{m}{n}
  \DeclareMathSymbol{\Gamma}{\mathalpha}{otone}{0}
  \DeclareMathSymbol{\Delta}{\mathalpha}{otone}{1}
  \DeclareMathSymbol{\Theta}{\mathalpha}{otone}{2}
  \DeclareMathSymbol{\Lambda}{\mathalpha}{otone}{3}
  \DeclareMathSymbol{\Xi}{\mathalpha}{otone}{4}
  \DeclareMathSymbol{\Pi}{\mathalpha}{otone}{5}
  \DeclareMathSymbol{\Sigma}{\mathalpha}{otone}{6}
  \DeclareMathSymbol{\Upsilon}{\mathalpha}{otone}{7}
  \DeclareMathSymbol{\Phi}{\mathalpha}{otone}{8}
  \DeclareMathSymbol{\Psi}{\mathalpha}{otone}{9}
  \DeclareMathSymbol{\Omega}{\mathalpha}{otone}{10}
  \DeclareSymbolFont{syms}{OML}{cmm}{m}{it}
  \DeclareMathSymbol{\partial}{\mathord}{syms}{"40}
  \DeclareMathAlphabet{\mathbold}{OML}{cmm}{b}{it}
  \DeclareSymbolFont{largesymbols}{OMX}{cmex}{m}{n}

\title{Sudden Attention Shifts on Wikipedia During the COVID-19 Crisis}
\author{
Manoel Horta Ribeiro,\textsuperscript{\rm 1}\thanks{\scriptsize These authors contributed equally. Also, this paper has been accepted at the 15th International Conference on Web and Social Media (ICWSM), please cite accordingly.}
Kristina Gligori\'c,\textsuperscript{\rm 1}$^*$
Maxime Peyrard,\textsuperscript{\rm 1}$^*$ \\
Florian Lemmerich,\textsuperscript{\rm 2}
Markus Strohmaier,\textsuperscript{\rm 3}
Robert West\textsuperscript{\rm 1}\\
}
\affiliations{

\textsuperscript{\rm 1}EPFL, 
\textsuperscript{\rm 2}University of Passau,
\textsuperscript{\rm 3}RWTH Aachen \& GESIS\\
\{manoel.hortaribeiro, kristina.gligoric, maxime.peyrard, robert.west\}@epfl.ch\\
florian.lemmerich@uni-passau.de, markus.strohmaier@cssh.rwth-aachen.de
}

\newcommand{\WP}{Wikipedia\xspace}
\newcommand{\covid}{COVID-19\xspace}

\begin{document}

\maketitle

\begin{abstract}
\vspace{-2mm}
We study how the COVID-19 pandemic, alongside the severe mobility restrictions that ensued, has impacted information access on Wikipedia, the world's largest online encyclopedia.
A longitudinal analysis that combines pageview statistics for 12 \WP language editions with mobility reports published by Apple and Google reveals massive shifts in the volume and nature of information seeking patterns during the pandemic.
Interestingly, while we observe a transient increase in Wikipedia's pageview volume following mobility restrictions, the nature of information sought was impacted more permanently.
These changes are most pronounced for language editions associated with countries where the most severe mobility restrictions were implemented.
We also find that articles belonging to different topics behaved differently; \eg, attention towards entertainment\hyp related topics is lingering and even increasing, while the interest in health- and biology\hyp related topics was either small or transient.
Our results highlight the utility of \WP for studying how the pandemic is affecting
people's
needs, interests, and concerns.

\end{abstract}

\section{Introduction}

\label{sec:intro}
The coronavirus disease 2019 (COVID-19) pandemic has led to the implementation of unprecedented non\hyp pharmaceutical interventions ranging from case isolation to national lockdowns~\cite{flaxman_report_2020}. 
These interventions, along with the disease itself, have created massive shifts in people’s lives. 
For instance, in mid-May 2020, more than a third of the global population was under lockdown~\cite{liverpool_covid19_2020}, and millions have since lost their jobs or have moved to work-from-home arrangements~\cite{hale2020oxford}.

\begin{figure}[ht]
\centering
\includegraphics[width=\linewidth]{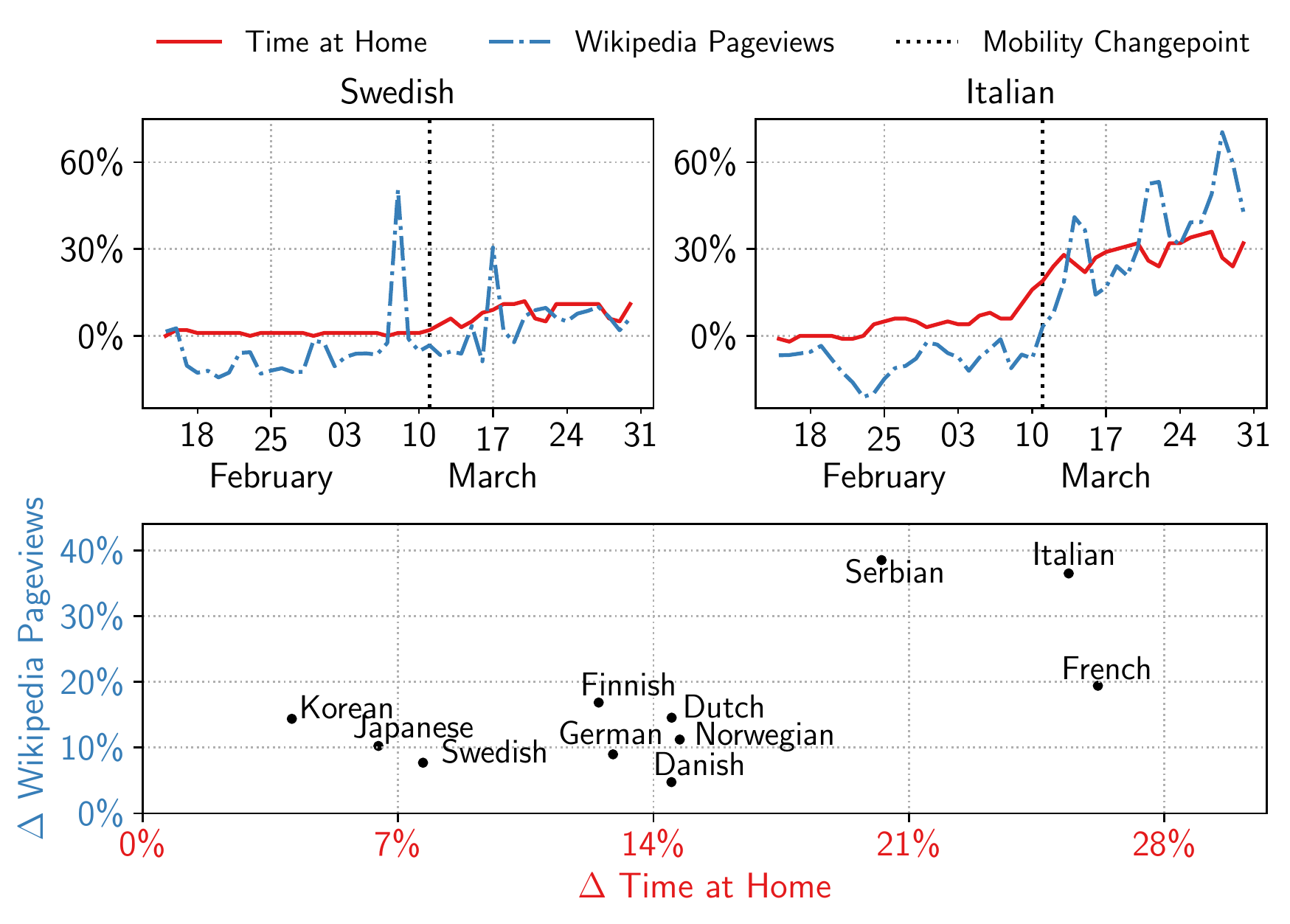}
\caption{Wikipedia access vs. mobility ---
Association between increase in time spent at home (from Google mobility reports; red) and increase in Wikipedia access volume (from \WP pageview statistics; blue),
both in terms of relative change over a five-week baseline period
in early 2020.
Top:
time series for Sweden\slash Swedish \WP and Italy\slash Italian \WP;
dotted vertical lines: changepoints in mobility time series.
Bottom:
summary for 11 of the 12 languages studied (excluding English);
$x$-axis: post-minus-pre\hyp changepoint difference in relative mobility change;
$y$-axis: post-minus-pre\hyp changepoint difference in total Wikipedia access volume
 (Pearson's $r = 0.63$, $p = 0.03$).
}
\label{fig:teaser}
\vspace{-5mm}
\end{figure}

This scenario has fueled an unprecedented effort to better understand the disease and its spread~\cite{cohen_unprecedented_2020}, as well as to develop clinical treatments and vaccines~\cite{who_172_2020,collins_accelerating_2020}.
Yet, recent work suggests that the impact of the pandemic transcends health-related issues~\cite{ryan_covid19_2020}. Thus, research should also identify how the pandemic has impacted human needs, interests, and concerns.
After all, many challenges of the \covid{} crisis did not originate directly from the virus, but from the social, economic, and psychological implications of the measures taken to prevent its spread~\cite{bonaccorsi2020economic}.

Quantifying the impact of \covid{} is particularly challenging due to its global nature: traditional survey-based studies face the difficulty of \emph{scale,} since participants ought to be spread throughout the globe, and of \emph{immediacy,} since the pandemic is an ongoing crisis where needs, interests, and concerns are dynamic~\cite{diaz2016online,salganik2019bit}.

Unlike most previous events that directly impacted so many lives around the world, the COVID-19 pandemic developed in a time of widespread Internet access. 
This digitization enables researchers to explore the impact of the pandemic across society by analyzing how it has impacted users' digital traces.
For instance, Wikipedia, the world's largest encyclopedia and one of the most visited sites on the Web, captures rich digital traces from readers and makes them publicly available in aggregated form.
Wikipedia is used hundreds of millions of times each day to address a wide spectrum of information needs, ranging from getting information to make personal decisions to reading up on what was discussed in the media~\cite{singer_why_2017}.

By analyzing Wikipedia access logs, we may thus study how the shifts in needs, interests, and concerns brought by \covid{} have affected the readership of a key information source on the Web,
while addressing
the challenges of \emph{scale} (since Wikipedia has versions for languages spoken all over the globe) and \emph{immediacy} (since the ``always-on'' nature of usage logs allows us to analyze information seeking patterns at any time).

\xhdr{Present work} This work aims to elucidate how information access patterns across 12 \WP languages editions have shifted during the current COVID-19 pandemic, in response both to the disease itself and to the massive mobility restrictions imposed by governments. 
More specifically,
we ask:
\begin{itemize}
    \item \textbf{RQ1}  How has the \covid{} crisis impacted information seeking behavior in Wikipedia? 
    \item \textbf{RQ2}  Have changes in information seeking behavior lasted after mobility restrictions were lifted?
\end{itemize}

We argue that a longitudinal understanding of information seeking through the lens of \WP is a starting point to measure the impact of the pandemic and its associated interventions on information seeking behavior. 
Examining the differences in the \emph{volume} (\ie, how much?) and in the \emph{nature} (\ie, what?) of information seeking may shed light not only on the needs, interests, behaviors, and concerns of people throughout the crisis, but also on how the pandemic is shaping society in the long run.
Also, observing what changes remained after mobility returned to normality can prove useful to disentangle effects due to the lockdowns from the drifts created by the pandemic itself. 

\xhdr{Methods and data}
Our analyses combine \WP access logs, mobility reports made available by Apple and Google, and information on the dates of non-pharmaceutical interventions.
We enrich the Wikipedia access logs in different ways, categorizing articles as \covid{}-related or not, attributing articles to a series of predefined ``topics'', and calculating how distant from normal is the distribution of views over articles.
Also, we calculate relevant points in time related to mobility change, capturing, for different countries, when mobility suddenly  decreased and when it returned  to normality.
Methodologically, drawing meaningful conclusions from the longitudinal \WP access logs is challenging due to the presence of trends and seasonalities.
We overcome these hurdles by performing careful difference\hyp in\hyp differences analyses in a regression framework.
We detail our methods and the data employed in \Secref{sec:matmet}.

\xhdr{Findings}
Our analysis of the number of pageviews over time (\Secref{sec:Shifts in overall pageview volume}) shows that there was a sharp increase in the \emph{volume} of \WP usage, by up to 40\%, following the sudden decrease in mobility induced by non\hyp pharmaceutical interventions (\cf\ \Figref{fig:teaser}).
This increase, however, was transient: our careful differences\hyp{}in\hyp{}differences study suggests that in the long run, there was either a decrease or no significant increase for 11 out of the 12 languages.

Considering not the volume, but the \emph{nature} of information seeking in \WP (\Secref{sec:Shifts from normality}), as measured through our ``distance from normality'' metric, we again observe the same patterns: a sharp increase following mobility restrictions, followed by a decrease after mobility returns to normality. 
Yet, here, our differences\hyp{}in\hyp{}differences study suggests that part of the increase remained, suggesting that the impact of the crisis on information seeking behavior persisted beyond mobility restrictions.

Lastly, we zoom in on what kinds of articles received more or less attention throughout the pandemic (\Secref{sec:Shifts in topic-specific pageview volume}). 
We find that some topics (\eg, \cpt{Biology}) saw transient increases or decreases in volume, while others saw persistent increases (\eg, \cpt{Video Games}) or decreases (\eg, \cpt{Sport}). Overall, surprisingly, many of the topics with persistent increases relate not to basic needs related to the pandemic, but to entertainment and self-actualization.

We conclude by discussing our findings in the light of the research questions and in the overall context of assessing the social impact of the pandemic  (\Secref{sec:discussion}).
We make available all the code and the data necessary to reproduce our results at: \url{www.github.com/epfl-dlab/wiki_pageviews_covid}.

\section{Related work}
\label{sec:rel}
We review related work that focused on the \covid{} infodemic, work that measured changes in user behavior during the pandemic, and, lastly, other research that studied \WP in the general context of crises and pandemics.

\subsection{The \covid{} Infodemic}

\covid{} was the first event of its magnitude to take place in the era of social media and of user-generated content. 
The important role that platforms such as Twitter and Facebook have in  today's society has prompted researchers to study
patterns of virality and of information sharing in social media during the pandemic.

\citet{kouzy2020coronavirus} studied the spread of \covid{}-related misinformation on Twitter by analyzing a sample of tweets ($n=673$), collected in February 2020, with trending hashtags and keywords related to the disease, finding a high prevalence of mis\hyp{} or unverifiable information. 
Other studies that followed have found similar and complementary results. \citet{yang2020prevalence} also found a high prevalence of low-credibility information, which is disproportionally spread by bots.
\citet{ahmed2020covid} identified and analyzed the drivers behind one of the main \covid{}\hyp{}related conspiracies, which postulates the spread of the infection to be related to the 5G standard for cellular networks. 
Analyzing data containing conspiracy-related hashtags, they found that a handful of users were driving the conspiratory content and that many of those using the hashtag were denouncing the conspiracy theory.


Whereas these studies detected and measured content that may impact behaviors and beliefs (\eg, conspiracy theories), we focus on the inverse question: \emph{Can we uncover changes in behavior via digital traces?}
We argue that both these directions are important to the overarching goal of understanding the social dynamics of the spread  of the virus \cite{depoux2020pandemic}, since to understand the impact of misinformation, one must be capable of measuring people's needs, interests, and concerns.

\subsection{The Web in Times of \covid{}}

Recent work has also more broadly characterized how \covid{} has altered people's online behavior, and how digital traces can be used to better understand the pandemic. 

A first and broader theme in this direction concerns how \covid{} has increased Internet traffic.
\citet{feldmann2020lockdown} showed that, as a result of COVID-19\hyp induced lockdowns, Internet traffic of residential users increased by 15--20\%. Traffic increases were observed in applications that people use when at home, such as Web conferencing, VPN, gaming, and messaging.
Results in the same direction were also found in survey-based studies analyzing Internet time~\cite{colley2020exercise}, and by a smaller-scale study measuring the increase in the stress on a campus network~\cite{favale2020campus}.

Second, and more related to the work at hand, are works leveraging digital traces to understand the impact of \covid{} on mental health, economy, society, and human needs~\cite{abay2020winners,gupta2020effects,tubadji2020narrative}.
We highlight two recent papers based on search data.
\citet{lin2020google} used Google search data on COVID-19\hyp specific keywords to predict the speed of the spread of the disease. 
They found, \eg, that searches for ``\textit{wash hands}'' are correlated with a lower spreading speed of the disease. 
\citet{suh2020population}  measured changes in human needs using Bing search logs.
They found that, for a variety of different ``need categories'', there was elevated increase in search activity, and that subcategories related to the most basic needs received the largest boost. 

The outlined related work is complementary to the study at hand. 
Having signals from multiple distinct digital traces such as network usage, social media activity, and search logs may allow stakeholders to paint a more comprehensive picture of how the pandemic has impacted society. 

\subsection{Wikipedia in Times of Crisis}

Finally, our work extends a rich literature studying behavioral changes of \WP readers during unexpected events, crises, and catastrophes \cite{garcia2017memory,zhang2019participation}.
In work that is most closely related, researchers used \WP pageviews in order to monitor and forecast diseases at a global scale \cite{generous2014global,10.1371/journal.pcbi.1004239}
and to study anxiety and information seeking about infectious diseases, such as influenza \cite{10.1371/journal.pcbi.1003581},
H1N1 \cite{Tausczik2012PublicAA},
and Zika \cite{zikacc}.

Resonating with existing work, our results highlight the integral role played by \WP in times of crisis and its usefulness to understand how these events impacted the behavior of its readers.
Moreover, it is worth noting that the difference\hyp{}in\hyp{}differences methodology we employ here may also be of use for future research aimed at understanding how a given crisis has impacted user behavior while accounting for seasonal effects.


\section{Materials and methods}
\label{sec:matmet}
\begin{table}[t]
    \centering
    \footnotesize
\begin{tabular}{lrrrr}
    \toprule
    Language &  \# articles &  2019 pageviews  & 2020 pageviews  \\
    \midrule
    English   &    6,047,509 &  83,566,105,101 &  51,911,047,562 \\
    Japanese  &    1,197,856 &  12,335,323,771 &   7,829,206,874 \\
    German    &    2,415,136 &  10,090,208,904 &   6,083,708,597 \\
    French    &    2,195,949 &   7,663,315,198 &   4,867,880,748 \\
    Italian   &    1,594,039 &   5,996,763,417 &   3,945,321,500  \\
    Dutch     &    2,003,807 &   1,678,509,656 &   1,010,487,782 \\
    Swedish   &    3,735,720 &   1,019,647,051 & 591,716,448 \\
    Korean    &     490,314 &    837,989,910 &    499,911,756  \\
    Finnish   &     481,854 &    665,104,786 &    399,430,747 \\
    Norwegian &     531,478 &    359,423,070 &    206,080,079 \\
    Danish    &     258,063 &    309,196,927 &    177,158,005  \\
    Serbian   &     632,128 &    274,704,611 &    206,425,442 \\
    \bottomrule
    \end{tabular}
    \caption{Basic dataset statistics: the number of articles in each \WP version, and the number pageviews received in 2019 (full year) and 2020 (January through July).}
    \label{tab:basic}
\end{table}
\subsection{Data}

The analyses of this paper combine information about the content and usage of \WP with information about the progression of the pandemic.

\xhdr{\WP}
We selected 12 languages that have \WP editions of various sizes.
When choosing language editions, we considered
(1)~the size of the edition,
(2)~whether the language was spoken in relatively few countries,
(3)~the kinds of non\hyp pharmaceutical interventions imposed in those countries.  
The overarching goal in selecting the languages was to have relevant language editions representing different attitudes towards the crisis, preferably from languages easily tied to one country.
For each language edition, we obtained publicly available pageview statistics,%
\footnote{\url{https://dumps.wikimedia.org/other/pageviews/}}
which specify how frequently each article was accessed daily between 1~January 2018 and 31~July 2020, aggregating both the desktop and the mobile versions of the site.
The languages selected as well as basic statistics for their \WP editions are listed in \Tabref{tab:basic}.

\xhdr{Enriching \WP articles}
To gain further insight into which kinds of articles people are reading, we enriched our pageview data with two additional sources.
First, we labeled each article with one of 57 topics based on the ORES \texttt{article topic} model.%
\footnote{\url{https://www.mediawiki.org/wiki/ORES/Articletopic}} We collected the topic predictions (one page can belong to multiple topics) from the model for each of the articles in English (the language for which the model was developed) and linked articles across the 12 languages studied.
Second, we determined for each article if it was related to the pandemic.\footnote{Based on the article list in \url{https://covid-data.wmflabs.org}. The number of COVID-19-related articles varied substantially across language editions; \eg, English had over 300, Swedish had only~9.}
Analyzing such articles separately allows us to disentangle the changes in \WP usage due to seeking information about the \covid{} pandemic from other, perhaps less obvious patterns.

\begin{figure}
\centering
\includegraphics[width=\linewidth]{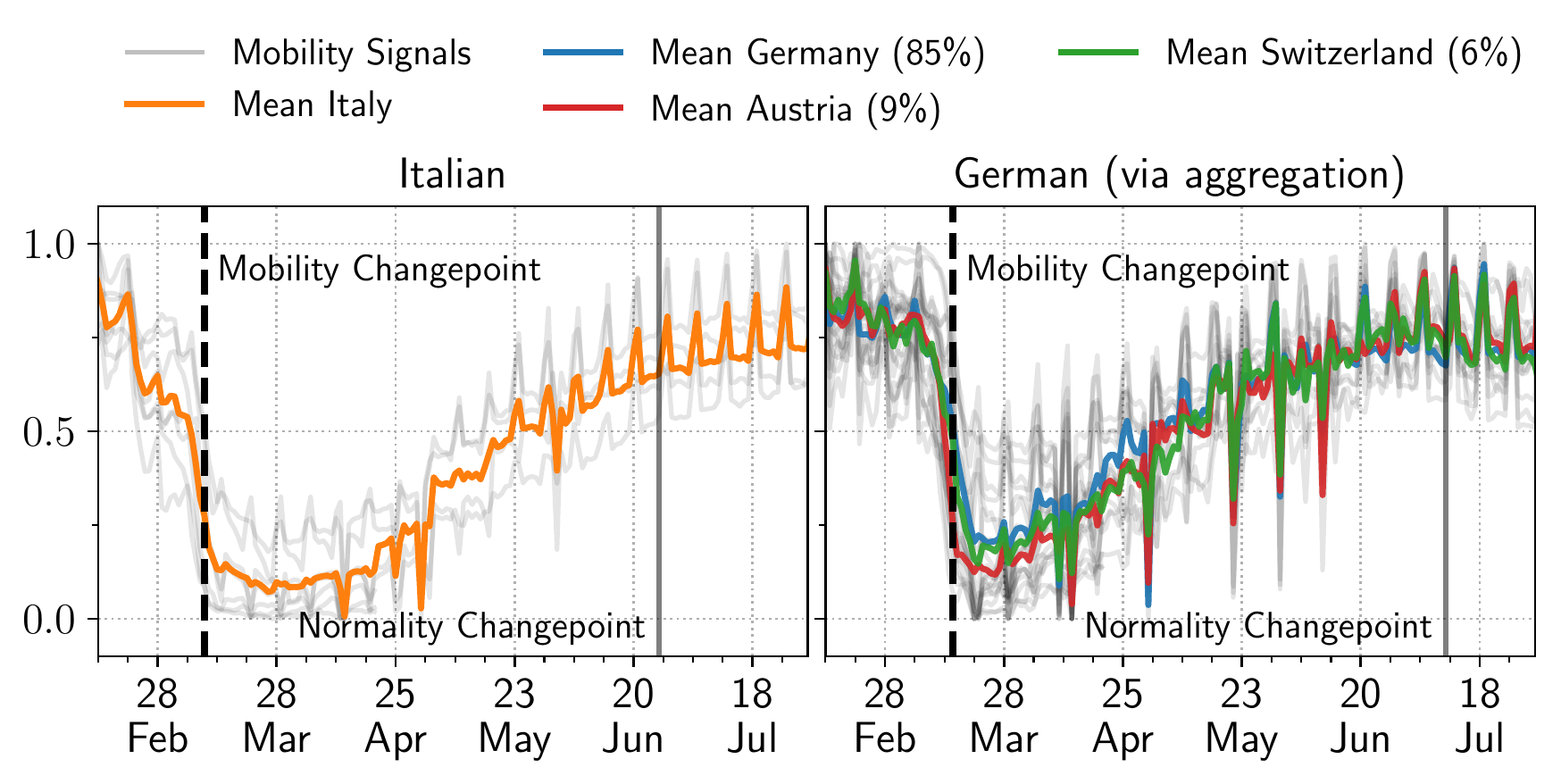}
\caption{
Robust detection of mobility and normality changepoints from Google and Apple mobility reports ---
The reports specify by what percentage the time spent at various location types has changed, compared to a baseline period in January 2020.
We use time series for all location types, with multiple changepoint detection algorithms and varying hyperparameters.
Given the abruptness of the decrease in most countries, different runs largely agree; we use the average as a robust changepoint.
For languages spoken in multiple languages, \eg, German (right), we average mobility time series from the main countries where the language is spoken (here three), weighted according to the proportion of native speakers (details in \Appref{sec:appendix}).}
\label{fig:changepoint_example}
\vspace{-2mm}
\end{figure}

\xhdr{Pandemic timeline}
Nine of the 12 languages are primarily spoken in a single country.
For each country, we used \WP to manually determine five days of particular interest in the context of \covid: first reported case, first death, ban of public events, school closure, lockdown.
Two problems with employing the aforementioned days of interest in statistical analyses are
(1)~that it is not guaranteed that they would impact movement patterns across different countries homogeneously (\eg, it could be that for some of the countries people stayed more at home even before the lockdown was enacted), and
(2)~that not all days of interest were observed for the countries of interest (\eg, in Sweden there has been no country-wide lockdown by 31~July 2020).

\xhdr{Mobility reports}
To address these limitations, we turn to daily mobility reports published by Apple and Google,%
\footnote{\url{https://www.(apple|google).com/covid19/mobility/}. We use country-wise mobility data from January to the end of July for all countries that have the languages chosen as official languages (see \Appref{sec:appendix}).}
which capture population\hyp wide movement patterns based on cellphone location signals.
The mobility reports specify, for each day, by what percentage the time spent in various location types (\eg, residential areas, workplaces, retail and recreation, \etc)\ differed from a pre\hyp pandemic baseline period in early 2020.
Both government\hyp mandated lockdowns as well as self-motivated social distancing measures  manifest themselves as sharp changes in the mobility time series, which we detect automatically using changepoint detection algorithms  (\Figref{fig:changepoint_example}, left; details in \Appref{sec:appendix}).
We henceforth refer to these points as \emph{mobility changepoints.} 
We use mobility changepoints as heuristic dates for when people started spending substantially more time in their homes. 
Unlike choosing one of the days of interest, this leads to a meaningful ``treatment'' across different countries.

Three of the 12 languages are spoken more widely than in one predominant country: English, German, and French.
As \WP pageview statistics are available only at the language level, not at the country level, we determined a mobility changepoint for these language editions by aggregating mobility reports for the countries in which the language is official (\Figref{fig:changepoint_example}, right; details in \Appref{sec:appendix}).

We emphasize that, since any \WP edition can be accessed from anywhere (barring censorship), the link between \WP language editions and countries is merely approximate, even for languages that are official in only a single country.
This should be kept in mind when interpreting our results, especially for the English edition, which is read widely across the globe.

Additionally, we calculated another important set of ``changepoints'', which do not represent abrupt changes in mobility patterns, but rather corresponds to the times when mobility returned to normal. 
We refer to these as \emph{normality changepoints} (also depicted in \Figref{fig:changepoint_example}).
These times were also calculated with mobility data, by identifying the time when the future average mobility remained within a $10\%$ band around the respective baseline level (defined as the pre-pandemic mobility levels by Google and Apple).
In cases of languages spoken in multiple countries, we maintained the same aggregation scheme as for \emph{mobility changepoints}.

\subsection{Difference\hyp{}in\hyp{}differences}
\label{subsec:diffsndiffs}


We now introduce our study design.
Difference\hyp{}in\hyp{}differences regression is a ``quasi\hyp experimental'' technique that mimics an experimental design with observational data by studying the effect of a treatment (\eg, mobility changes) on a treatment group \vs\ a control group~\cite{angrist1995identification}.
The specific regression models are discussed as we present results, whereas here we focus on explaining and justifying our setup, illustrated on the left-hand side of \Figref{fig:diffndiff}.

The difference\hyp in\hyp differences method aims at separating the true treatment effect from simultaneous (\eg, seasonal) changes that would have occurred even without the treatment.
To do so, we calculate the post\hyp minus\hyp pre\hyp treatment difference (in 2020) and compare it to the difference between the corresponding time periods in the previous year (2019).
Subtracting the 2019 difference from the 2020 difference (yielding the ``difference in differences'') thus removes changes that would have occurred even without the intervention (assuming 2020 would otherwise have looked like 2019) and gives a better estimate of the treatment effect.


In the first scenario, \Figref{fig:diffndiff}(a), we compare the difference in activity five weeks \emph{before} \vs\ five weeks \emph{after the mobility changepoint.}
The idea here is to capture what changes were introduced by the sudden halts in human mobility induced by non\hyp phamaceutical interventions.
In the second scenario, \Figref{fig:diffndiff}(b), we compare the difference five weeks \emph{after the mobility changepoint} \vs\ five weeks \emph{after the normality changepoint.} 
Here, we capture how information seeking patterns changed as mobility restrictions waned.
The third scenario, \Figref{fig:diffndiff}(c),  compares the difference  in activity between two scenarios where mobility was similar in 2020: five weeks \emph{before the mobility changepoint} and five weeks \emph{after the normality changepoint.}
Lastly the fourth scenario, \Figref{fig:diffndiff}(d),  is an adaptation of the first for events other than mobility (\eg, school closure or official lockdown declaration).

Notice that for some languages there were less than five weeks in the data after the normality changepoint. In these cases we used all data up to 31 July 2020 (and the same period of 2019).

\subsection{Distance from Normality}
\label{sec:distnorm}

We introduce a notion of ``distance from normality'' as follows. 
On each day, the pageviews in a given language edition form a distribution over articles, characterizing how users' attention was distributed.
We represent each daily distribution as an ``attention vector'' with one entry per article and entries summing to~1. 

With over 6~million Wikipedia articles, many of which are rarely visited, attention vectors are large and noisy. Therefore, we first applied principal component analysis (PCA) in order to project attention vectors into a low\hyp dimensional subspace.
In the subspace, two attention vectors are naturally compared via their Euclidean distance.
 
The notion of ``normal'' attention is captured by the average attention vector over all days of 2019, \ie, well before the pandemic;
and for each subsequent day, the distance from normality is given by the Euclidean distance of that day's attention vector from the average attention vector.
Notice that we calculated the attention vector separately for each \WP language edition using all its articles and then performed the dimensionality reduction.

We use this metric in addition to the aforementioned topics to understand overall changes in the information seeking patterns of \WP users.

\begin{figure*}[t]
\centering
\includegraphics[width=\textwidth]{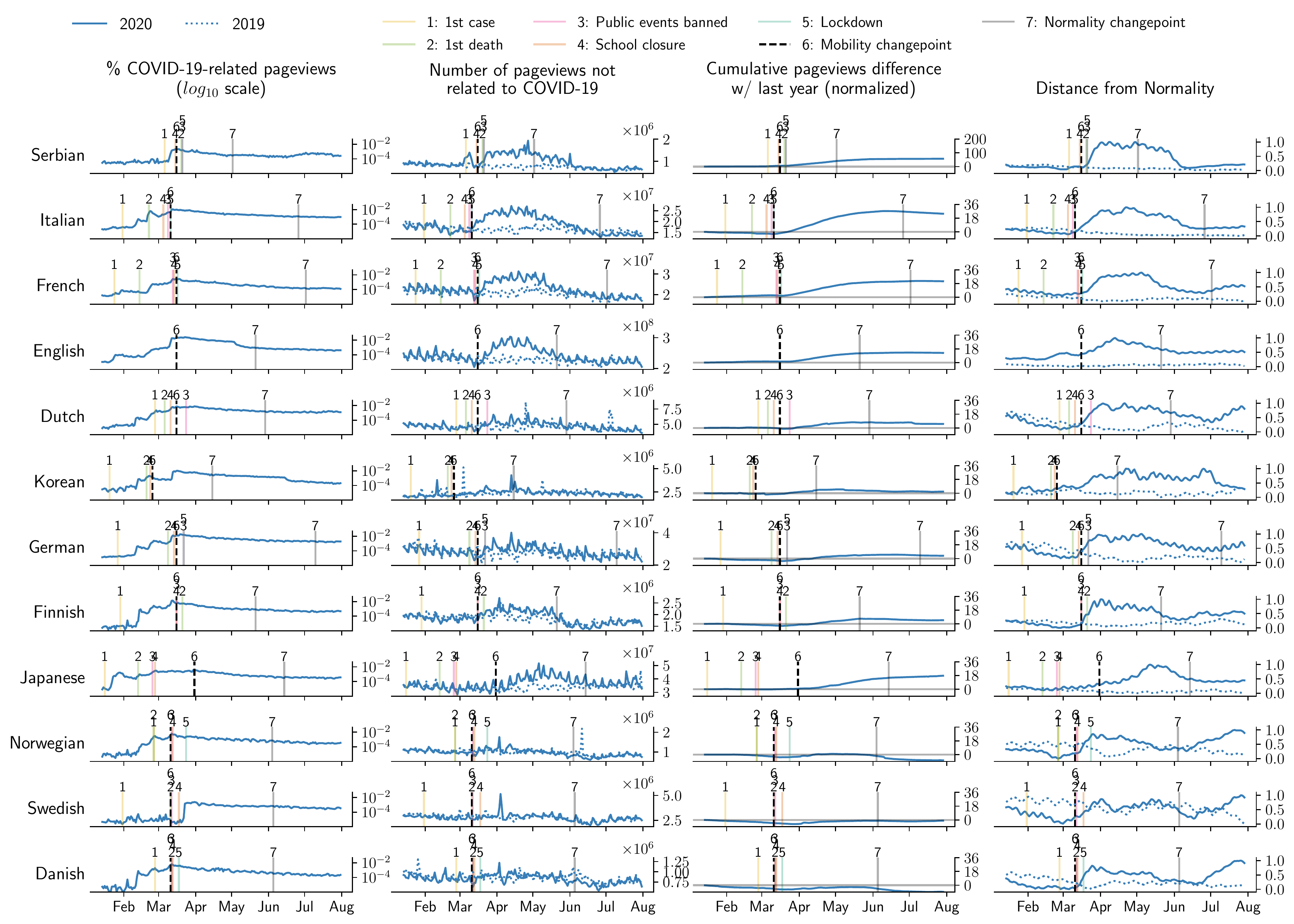}
\caption{
Evolution of Wikipedia access for 12 language ---
Column~1:
percentage of pageviews spent on COVID-19\hyp related pages (logarithmic scale).
Column~2:
total number of pageviews for all other pages (linear scales);
solid: 14~January to 31~July 2020;
dotted: corresponding period in 2019.
Column~3:
cumulative difference in total number of pageviews between study period of 2020 and corresponding period of 2019,
reported in terms of multiples of average daily number of pageviews in 2019
($y$-axes identical across languages, except for Serbian, where cumulative volume is much higher).
Column~4:
``distance from normality'' with respect to topical attention (\cf \Secref{sec:Shifts from normality}) for 2020 (solid) and 2019 (dotted).
All plots show, as vertical lines, mobility changepoints and  pandemic\hyp related events.
}
\label{fig:pageviews}
\end{figure*}

\section{Results}
\label{sec:res}
\subsection{Shifts in Overall Pageview Volume}
\label{sec:Shifts in overall pageview volume}

We begin by examining the \emph{volume} of information seeking throughout the pandemic. 
To do so, we leverage the count of pageviews across all articles for each of the 12 languages.
To distinguish between an increase in information seeking about \covid{}\hyp related articles \vs the remainder of \WP, we analyze the two sets of articles separately.

\xhdr{Pageview trends in \covid{}\hyp related articles}
\WP has been shown to be an accurate and up-to-date source of \covid{}\hyp related information,%
\footnote{\url{https://wikimediafoundation.org/covid19/}}
and we start by investigating how heavily this information was accessed by users.
\Figref{fig:pageviews} (column~1) tracks the popularity of \covid{}\hyp related articles (as a fraction of all pageviews) in all 12 languages over the course of the pandemic, from 14~January (the earliest time for which mobility reports are available) to 31~July 2020.
Vertical lines in the figure represent the dates associated with noteworthy events (\eg, first case), as well as non-pharmaceutical interventions (\eg, lockdowns).

In nearly all languages, the share of \covid{}\hyp related pageviews increased up to the mobility changepoint (dashed vertical line), from where their daily pageviews tended to slightly decrease.%
\footnote{
Some languages saw extreme upticks on certain days.
Most of them can be linked to the creation of important \covid{}\hyp related articles;
\eg, the Swedish article
\cpt{Coronavirusutbrottet 2020 i Sverige} (English: \cpt{COVID-19 pandemic in Sweden}) was created on 22~March 2020 and immediately received wide attention.}
The eventual return to normal levels of mobility did not seem to have any sharp impact on the pageview share of these articles.
We emphasize that these time series are plotted on logarithmic $y$-scales, such that a linear slope in the plots, even if small, corresponds to an exponential rate of change.

\covid{}\hyp related articles were generally among the most popular during the period of study; \eg, 12 of the 15 most accessed articles in the English version were related to the pandemic.
In some languages, the fraction of pageviews going to \covid{}\hyp related articles surpassed 2\% on some days, a considerable share of \WP's overall volume, considering that all \WP editions considered here have over half a million articles (\Tabref{tab:basic}).

\xhdr{Pageview trends in non-\covid{}\hyp related articles}
Next, we focus on \WP articles \emph{not} related to \covid{}, the vast majority.
Their popularity during the period of study, in terms of the daily total number of pageviews, is shown as solid lines in \Figref{fig:pageviews} (column~2; linear $y$-scales).
The dotted lines correspond to the same period in 2019, precisely one year earlier.%
\footnote{
Since \WP access volumes tend to follow a weekly periodicity, our alignment was manually adjusted to ascertain that matched days correspond to the same day of the week.
}
Notice that, in this plot, the  $y$-scales vary per language edition, since their sizes differ substantially.

A clear pattern emerges in many language editions.
Access volumes in 2020 closely mirrored those in 2019 up to the mobility changepoint associated with the respective language.
Thereafter, the access volumes of 2020 began to rise up and above those of 2019 for nearly all languages.
The trend was particularly strong for languages spoken in the countries with the most severe lockdown measures (\eg, Serbia, Italy, France), and weakest for languages spoken where mobility-related interventions were weaker than elsewhere (\eg, Japan, Scandinavia).
Finally, once normality was approximately restored (as indicated by the normality changepoint), we see that the pageview volume again aligned itself with the quantities observed in the previous year. 
Again, for countries with weaker lockdowns, the ``back to normality'' effect is less noticeable since the pageview counts never deviated substantially from the baseline pattern to begin with.

We further explore this relationship in \Figref{fig:pageviews} (column~3) via cumulative plots, where the daily 2020-minus-2019 difference is accumulated from 14~January onward.
Since languages have vastly different access volumes overall (\cf\ column~2), and to have a common scale across languages, we express the cumulative difference in terms of multiples of the average daily pageview volume attained in 2019 by the respective language.
The increase is dramatic for some languages; \eg, the Serbian \WP edition experienced about 57 days' worth of surplus access volume between mid-March and late July 2020, corresponding roughly to a tripling in access volume, compared to the previous year. 
Strong effects are also observed for
Italian (27 surplus days at its peak),
French (21 days),
Japanese (17 days),
English (13 days),
Dutch (7 days), and
Finnish (7 days).
As mobility returned to normality, the cumulative gain stabilized (\eg, Serbian, French) or even decreased (\eg, Dutch, Italian).

\begin{figure*}[t]
\centering
\includegraphics[width=\linewidth]{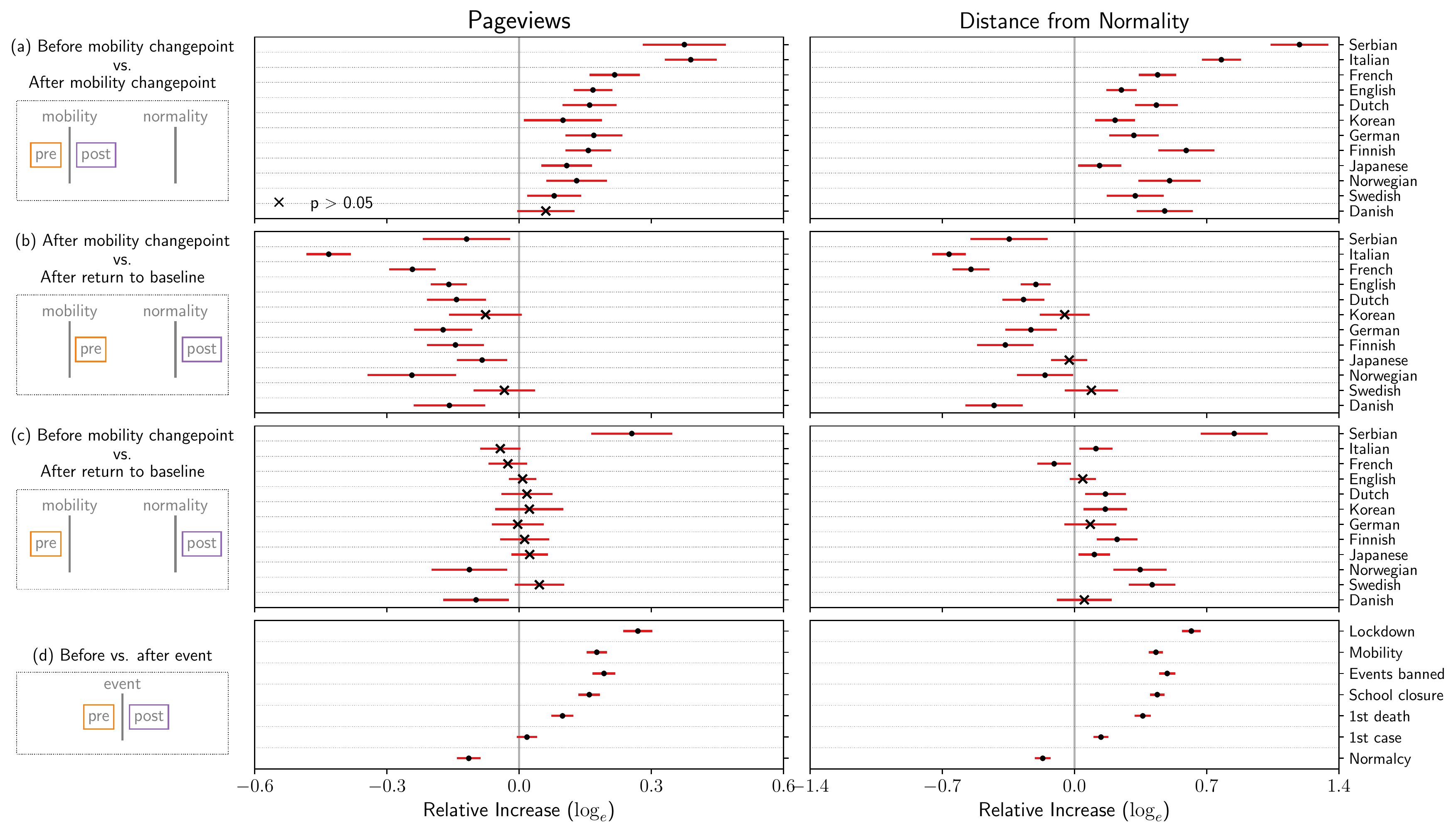}
\caption{
Estimated effects of restricted mobility in pageviews and in attention --- For the 12 studied language editions, we depict the results of difference\hyp in\hyp differences for estimating effects of changepoints non-pharmaceutical intervention  in the total number of pageviews (on the left) and in the distance from normality metric (on the right).
Effects are shown for different diffs-in-diffs setups illustrated in diagrams in the left-hand side (see \Secref{subsec:diffsndiffs} for details).
Error bars represent 95\% CIs approximated as 2 standard errors.
Dependent variables are used in logarithmic form, such that exponentiated coefficients capture the multiplicative increase due to the treatment.
($R^2>0.95$ for all models.)
}
\label{fig:diffndiff}
\end{figure*}

Note that eight of the 12 languages (Italian, Japanese, Finnish, Dutch, German, Korean, Danish, Serbian) initially ran a deficit, with \WP being visited \emph{less} in 2020 compared to the corresponding days in 2019 (reflected as negative values in the plots of column~3 of \Figref{fig:pageviews}), but all except three languages (Norwegian, Swedish, Danish) recovered and eventually ran a surplus by the end of the study period.

\xhdr{Difference-in-differences regression}
In order to go beyond visual inspection and to precisely quantify the shifts in pageview volume, we take a regression\hyp based difference\hyp in\hyp differences approach, as described in \Secref{subsec:diffsndiffs}.

In this setup, we consider, for each language, a time window of 10 weeks (70 days) split around either the mobility or the normality changepoint in 2020, as well as the corresponding time window in 2019.
Each of these 140 days contributes one data point per language,
for a total of $140 \times 12 =$ 1,680 data points.
As the dependent variable $y$, we use the logarithm of the number of pageviews,
and as independent variables, the following three factors:
$\mathtt{year}$ (2019 or 2020),
$\mathtt{period}$ (before or after calendar day of mobility changepoint),
$\mathtt{language}$.
We now model $y$ as a linear function of these three factors and all their two- and three-way interactions.
In R formula notation,
\begin{equation}
    y \sim \mathtt{year * period * language}.
    \label{eqn:formula_overall_volume}
\end{equation}
Here, $\mathtt{a * b}$ is shorthand notation for $\mathtt{a + b + a:b}$, where in turn $\mathtt{a:b}$ stands for the interaction of $\mathtt{a}$ and $\mathtt{b}$.

Pageview volumes were considered in logarithmic form for two reasons:
first, because raw pageview counts are far from normally distributed, with numerous large outliers,
and second, because the logarithm makes the model multiplicative, thus implicitly normalizing the estimated effects and making it possible to compare languages with different pageview volumes:
if $b$ is the coefficient of the three-way interaction $\mathtt{year : period : language}$, then $e^b$ captures the multiplicative factor by which pageview volumes increased when mobility dropped,
after accounting for differences stemming from the year alone or the period alone,%
\footnote{
For instance (\cf\ \Figref{fig:pageviews}, column~2), Swedish pre-mid-March pageviews were consistently lower in 2020 than in 2019; and Finnish post-mid-March pageviews were higher than pre-mid-March pageviews even in 2019, without the pandemic.
}
which are already captured by the coefficients of $\mathtt{year:language}$ and $\mathtt{period:language}$, respectively.

The estimated logarithmic effects are plotted for all languages in \Figref{fig:diffndiff}(a) (column~1).
The results confirm the effects observed visually in \Figref{fig:pageviews} (column~3);
\eg, the logarithmic pre-\vs{}-post mobility changepoint effect on the Italian version is around 0.38 (corresponding to an increase in pageviews to $e^{0.38} \approx 146\%$).
For language editions where the effect was less pronounced in the visual analysis, we also find smaller (\eg, for Swedish the increase was to approximately 108\%) or insignificant (\eg, Danish) effects.
Another set of interesting cases are languages for which the increase in pageviews was substantial but more gradual, \eg, Japanese. Although Japanese saw a larger cumulative pageview difference than English (\Figref{fig:pageviews}~column 3), here its effect is smaller (0.10 for Japanese \vs 0.16 for English).

In a similar fashion, \Figref{fig:diffndiff}(b) (column~1) shows the same analysis, but now comparing the five-week period \emph{after the mobility changepoint} with the (up-to-)five-week period \textit{after the normality change point}.
Interestingly, the results of this analysis are, to a large extent, diametrically opposite of those of the above analysis. 
Whereas there (around the mobility changepoint), Serbian and Italian saw the biggest increases in pageview volume, here (around the normality changepoint), they saw the biggest decreases.
For instance, the logarithmic effect is $-0.43$ for Italian (a multiplicative decrease to $e^{-0.43} \approx 64\%$ of the original value).

In \Figref{fig:diffndiff}(c) (column~1), we fit yet another model that shares the pre-intervention with the setup in \Figref{fig:diffndiff}(a) and the post-intervention period with \Figref{fig:diffndiff}(b).
Its purpose is to determine long-term effects, \ie,
whether there were any changes between the periods before the mobility changepoint and after the normality changepoint.
From \Figref{fig:diffndiff}(c) (column~1), we observe that most languages saw no long-term effects, whereas two (Norwegian and Danish) saw a significant negative long-term effect, and only Italian saw a significant positive long-term effect.

Lastly, in the fourth and last setup, we fit a slightly different model:
\begin{equation}
    y \sim \mathtt{year * period + language},
    \label{eqn:slightly_different_model}
\end{equation}
where the dependent variable $y$ is again the logarithm of the number of pageviews, $\mathtt{language}$ now merely serves as a language\hyp specific baseline, and $\mathtt{year : period}$ is the only interaction term which, captures the estimated effect in a language\hyp independent way.
We fit this model not only for the case where $\mathtt{period}$ is defined by the mobility changepoint, but also where it is defined by the five other pandemic-related events.
This way, we may compare effect sizes for the various events when considered as ``treatments''.

The estimated logarithmic effects are plotted for each event in \Figref{fig:diffndiff}(d) (column 1).
We find that events that are more tightly related to decreased mobility are associated with the largest increases in pageviews.
For example,
the pre-\vs{}-post first-death effect is only around 0.09 ($e^{0.10}\approx110\%$);
for school closure, it is around 0.16 ($e^{0.16}\approx117\%$);
and for the actual lockdown, it is largest, at 0.27 ($e^{0.27}\approx131\%$).
Moreover, as expected from the  previous results, we also find that the effect around the normality changepoint is significantly negative, at $-0.11$ ($e^{-0.11}\approx89\%$).

This corroborates the previous results on mobility and \WP access from yet another angle.
Earlier, in \Figref{fig:pageviews} and \Figref{fig:diffndiff}(a), we showed that languages spoken in countries with a stricter lockdown saw a larger pageview increase,
whereas here we showed that not only does this hold when considering all languages together, but also that events associated with a mobility decrease are significantly more associated with an increase in pageviews than other pandemic\hyp related events.

\begin{figure*}[t]
    \begin{minipage}[t]{\textwidth}
        \centering
        \includegraphics[width=\textwidth]{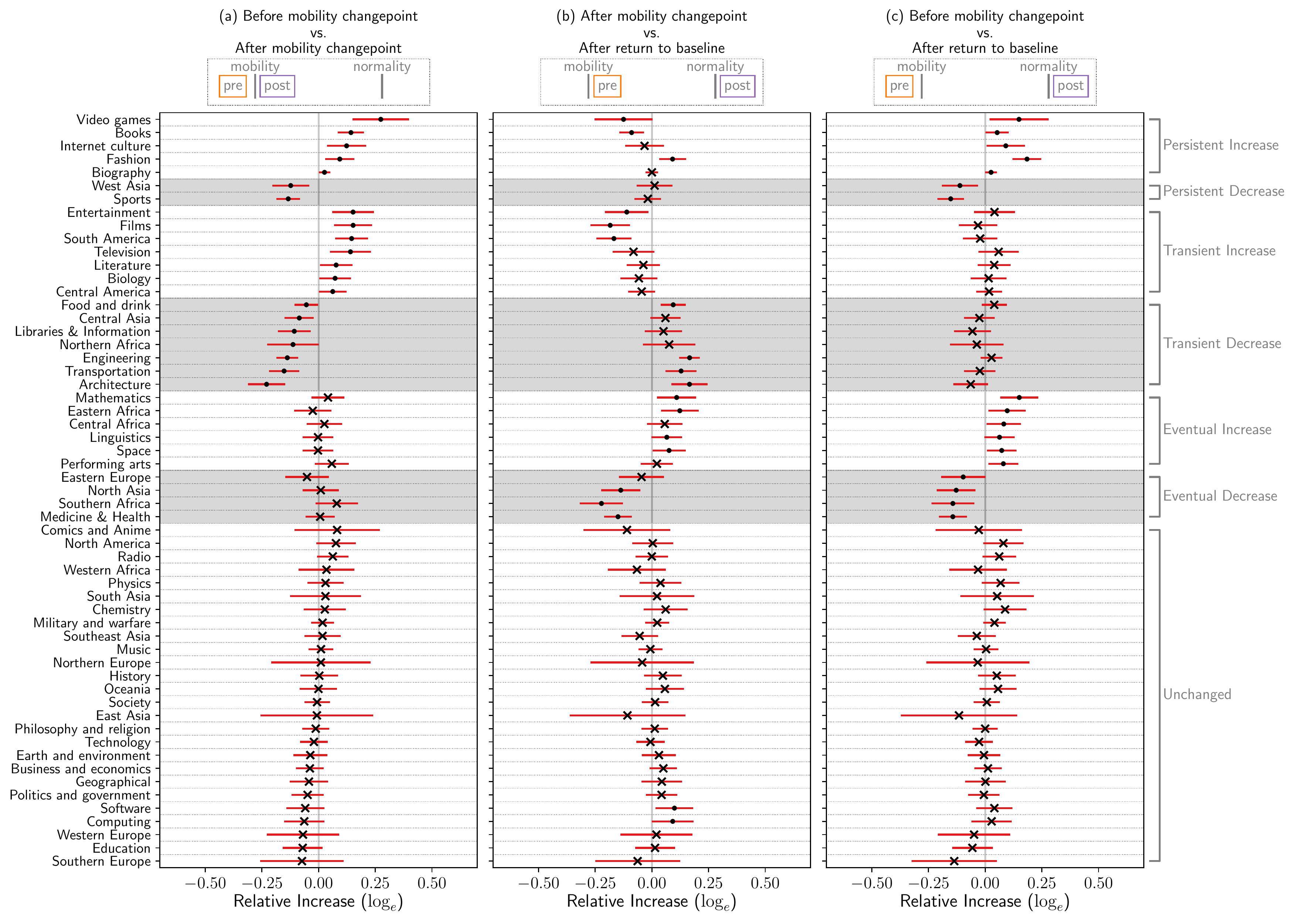} 
    \end{minipage}
    
    \caption{
    Shifts in topic-specific pageview volumes ---
    Effect of mobility decrease (and eventual return to normalcy) on relative pageview share of 57 topics, estimated via difference\hyp in\hyp differences regression, pooled across 12 languages. Again we use three different setups in order to capture how things changed through different moments of the pandemic (see \Secref{subsec:diffsndiffs} for details). 
    Topics are grouped according to their behaved given changes in human mobility, as shown in the brackets in the right-handside.
    Error bars represent 95\% CIs. 
    ($R^2 > 0.85$ for all models.)
}
    \label{fig:topics}
\end{figure*}

\subsection{Shifts in Information Seeking Patterns}
\label{sec:Shifts from normality}

So far, we observed an overall increase in pageviews following mobility restrictions. 
Then, once mobility went back to normal, so did pageviews.
Next, we explore whether these shifts affected pageviews evenly across articles or whether users' attention shifted from certain topics to others.


To gain a better understanding, we analyze the temporal trends for the distance\hyp from\hyp normality metric (defined in \Secref{sec:distnorm}) and perform a difference\hyp in\hyp differences analysis in the same setup as in \Secref{sec:Shifts in overall pageview volume}.

\xhdr{Distance from normality} 
Distances from normality are plotted as time series in \Figref{fig:pageviews} (column~4).
We also plot a baseline time series computed on data from exactly one year earlier (dotted line), where normality is defined as the average attention vector of 2018, and distance from normality was computed for mid-January to the end of September 2019.
We observe that the (dotted) 2019 baseline curves stay flat around the mobility changepoints.
If additional pageviews gained after the changepoints mirrored the average attention distribution, one would expect equally flat curves for 2020.
Quite on the contrary, however, the curves for 2020 increase sharply at the mobility changepoints, a clear indication of large topical shifts in overall attention.
Interestingly, we find that these changes occur even for languages for which there was no substantial pageview increase, such as Danish, Swedish, and Norwegian.
This suggests that, although these languages were not reading more, their information seeking patterns still differed overall.

As mobility returned to normality, we find that \WP readers' attention also shifted back towards normality. 
That is, the distance from normality became smaller after the effect of mobility restrictions waned.
Yet, it is important to notice that the level at which the distance from normality settled thereafter was generally larger than in the previous year (\eg, 0.3 \vs 0.03 for Italian).
Also, for some language editions (Dutch and Scandinavian languages), we see a second sudden increase in the metric around mid-July, for which we found no obvious explanation.

\xhdr{Difference-in-differences regression} 
To quantify the attention shifts more objectively, we perform a difference\hyp in\hyp differences regression analogous to the one of \Secref{sec:Shifts in overall pageview volume} (\Eqnref{eqn:formula_overall_volume}), but this time with the outcome $y$ being the logarithm of the distance from normality, rather than of the pageview volume.
We again employ the setups previously described and show the interaction terms on the right-hand side of \Figref{fig:diffndiff}.
Overall, in the two first setups (\Figref{fig:diffndiff}(a-b), column~2), we see results similar to those obtained in the pageview analysis.
For all languages, there was a sharp increase in the distance from normality when comparing the periods before \vs\ after the mobility changepoint.
Comparing the period after the mobility changepoint with the period after the normality changepoint, we observe a significant decrease.
Results are also similar to what we observed for pageviews in the fourth scenario,  (\Figref{fig:diffndiff}(d)), where we compare ``before \vs after'' for several pandemic-related events. 
Here again, we see a gradation, where the regressions for events more closely associated with mobility decrease (\eg, lockdowns) are associated with stronger effects.

Interestingly, the picture in the third scenario (\Figref{fig:diffndiff}(c)) differs from the pageview analysis: here the distance from normality remains higher than baseline for eight of the 12 languages. This scenario compares the period before the mobility changepoint with the period after the normality changepoint. This indicates that attention shifted massively during the mobility restrictions but settled in a new state after mobility went back to normal. 

This suggests that a portion of the effect was transient and clearly associated with the mobility decrease, but that there was also another part that lingered months after mobility decreases waned.
A plausible explanation for this phenomenon could be that the societal impact of the pandemic and its associated non\hyp pharmaceutical interventions have altered the interests, needs, and concerns of people for good.

\subsection{Shifts in Topic-specific Pageview Volume}
\label{sec:Shifts in topic-specific pageview volume}

So far, we observed a massive increase in pageview volume following the sudden mobility change induced by \covid{}-related interventions, which was not explained by a simple proportional increase according to the prior attention distribution (\Secref{sec:Shifts in overall pageview volume}). 
Rather, the increase in pageview volume was accompanied by major shifts in attention (\Secref{sec:Shifts from normality}).
Moreover, we have also found that, although the shifts in pageview volume were reverted once mobility restrictions waned, this was not the same for the shifts in attention: information seeking patterns seem to have been impacted in the long run.
Next, we investigate the shift further, with the goal of identifying which topics gained, and which lost, attention, and also which of these changes remained after the mobility reverted to normality.

Again, we account for trends and seasonality by formulating difference\hyp in\hyp differences regression models analogous to that of \Eqnref{eqn:formula_overall_volume}, but we now use a model without a language term and with an added topic term (for notation, \cf\ \Eqnref{eqn:formula_overall_volume}):
\begin{equation}
    y \sim \mathtt{year * period * topic}.
    \label{eqn:formula_topics_no_lang}
\end{equation}
We code the 57 topics as 56 dummy variables, with one arbitrary topic serving as a baseline topic.
This amounts to 95,760 data points, one for each combination of year, calendar day (35 days before as well as after), topic, and language.
The outcome variable $y$ of interest is the logarithm of the fraction of pageviews going to articles of the respective topic on the respective day. We again fit models for different setups (here we consider only the first three scenarios discussed in \Secref{subsec:diffsndiffs}).
Summing the coefficient of $\mathtt{year : period : topic}$ with the coefficient of $\mathtt{year : period}$ (corresponding to the baseline topic) reveals changes in topical interest around the changepoints of interest.
Given the three distinct difference\hyp{}in\hyp{}differences estimations, we are able to categorize different topics according to how they behaved given changes in human mobility: Did they change? Did these changes endure once mobility went back to normal?
The 57 topics are annotated with these different ``behavioral groups'' on the right-hand side of \Figref{fig:topics}. 

In the first two sectors of the figure (the first two colored blocks), we have topics for which a persistent increase or decrease was observed. These topics experienced an increase following the mobility changepoint, as shown in column (a), which was sustained once the pandemic ended, as shown in column (c).
Among the topics for which there was a persistent \textit{increase} in pageview volumes, we see entertainment\hyp related subjects such as \cpt{Video Games}, \cpt{Books}, and \cpt{Internet culture}.
Meanwhile, topics that experienced a persistent \emph{decrease} include a geographical location (\eg, \cpt{West Asia}) as well as \cpt{Sports}, a topic clearly related to outdoor activities.
These findings suggest that people shifted information seeking needs from outdoor (\eg, \cpt{Sports}) to indoor activities (\eg, \cpt{Video games}).

In the third and fourth sections of the figure, we have topics for which the increase or decrease in pageview volume was transient.
A noteworthy topic here is \cpt{Biology}, which may have received a boost given the sudden importance of all things virus-related, but that eventually, as people got used to the pandemic, returned to the normal levels of volume compared to the previous year. After the mobility changepoint, as shown in column~(a),  pages related to biology had an increase to around 109\%, followed by a decrease to around 89\% after things went back to normal, as shown in column~(b). 
Overall, comparing the five-week period before the mobility changepoint with the five-week period after the normality changepoint, the effect is not significant.

Lastly, the last five sections of the figure contain either topics that observed a ``reversed'' increase or decrease,\footnote{That is, they changed in one direction with the decrease in mobility, but then experienced a ``bounce back'' after mobility went back to normal.} shown in the sixth and seventh sections, or that experienced no change following the mobility changepoint at all, in the eighth, nineth and tenth sections. 
For the latter kind, some topics also saw an eventual increase or decrease in the difference\hyp in\hyp differences scenarios depicted in columns (b) or (c).
Of these last sections, a remarkable topic is \cpt{Medicine \& Health}, which presents a substantial decrease overall, as shown in column (c), but which did not experience an increase following the mobility changepoint, as shown in column (a).
We further investigated the time series for this topic and found that this happened because the shift in interest for the topic happened \emph{before} the mobility changepoint for many countries.
This is aligned with previous suggestions that the ``panic'' of the disease preceded its actual spread~\cite{depoux2020pandemic}.

All in all, our results suggest that, while \covid{}-related topics saw transient increases (\eg, \cpt{Biology}) or no increase at all (\eg, \cpt{Medicine \& Health}) following mobility restrictions, several topics related to entertainment saw long-lasting increases. Similarly, \cpt{Sports}, a topic related to the outdoors, saw persistent decreases. 
Importantly, although mobility returned to normal in many places, the pandemic is far from ending.
It will be interesting to see whether these shifts are medium-term changes that will revert once the pandemic is over or long-term changes that will remain for years to come.

\section{Discussion and conclusions}
\label{sec:discussion}
In summary, our findings suggest that the sharp decrease in human mobility induced by \covid{}-related non\hyp pharmaceutical interventions has boosted the \emph{volume} of information seeking on \WP and has changed the \emph{nature} of the information sought, even considering non-\covid{}-related articles (RQ1). 
Once mobility returned to its normal levels, the volume of information seeking also returned to its prior levels, but the kinds of content sought did not (RQ2).
Lastly, touching upon both research questions, we zoom in on what articles received more or less attention during this period, finding that some topics, \eg, \cpt{Biology}, saw transient increases or decreases in volume, while others saw persistent increases (\eg, \cpt{Video Games}) or decreases (\eg, \cpt{Sport}).

These findings may be help in  characterizing the societal impact of the pandemic.
First, the trends in the \emph{volume} of pageviews may themselves be interpreted as a trace of the social impact of the pandemic. 
The increase in information seeking that closely followed changes in mobility patterns suggests that, once we are deprived of activities halted by non-pharmaceutical interventions, we resort even more to information seeking on the Web, an activity related to entertainment and self-actualization.

Second, our analysis of the \emph{nature} of what the information sought suggests that the societal impact of the crisis has persisted beyond the associated mobility restrictions.
This acts as evidence that long-term changes induced by the pandemic are not only geopolitical or economic in nature, but also related to human needs, interests, and concerns.

Third, our results suggest that most of the topics that experienced persistent increases (\eg, \cpt{Video Games} and \cpt{Internet culture}) are not related to needs often framed as ``basic'', such as safety and physiological needs~\cite{maslow1943theory}.
Although this does not imply that the latter were not increased,%
\footnote{And indeed previous work has found evidence of increases on queries related to such basic needs~\cite{suh2020population}.}
it nevertheless reinforces that there was an increase in people's need for entertainment and arguably self-actualization.

An alternate angle in which we can frame the aforementioned discussion is to consider not how \WP logs shed light on the societal impact of the pandemic, but how this specific event sheds light on \WP's role in times of crisis.
While previous work has highlighted how \WP is able to provide crucial information to  millions of people during epidemics~\cite{10.1371/journal.pcbi.1003581}, we show that things do not stop there:
\WP can also play the role of providing information related to non-essential needs, and browsing \WP becomes more widespread once other activities are hindered by an unexpected crisis.

\xhdr{Limitations}
In this paper, we performed a careful analysis of access patterns on \WP.
As our findings are observational, we cannot conclusively establish causal effects of pandemic\hyp related events on user behavior.
Indeed, lockdown measures and the ensuing decrease in mobility coincided with numerous other life changes inflicted by the pandemic, and any of these concomitant changes---rather than the increase in time spent at home---might in principle be the true cause of the changes in \WP access patterns.
For instance, one might argue that an increased concern about health issues might drive people to access more encyclopedic information.
Nevertheless, our work presents circumstantial indications that there is indeed a causal link between mobility and the volume and nature of information seeking, as
we find that \WP\hyp{}derived signals follow mobility patterns closely (\Figref{fig:diffndiff}a-b analyzed jointly) and that there exists a dose--response relationship between mobility interventions and \WP\hyp{}derived signals (\Figref{fig:diffndiff}d).

Another limitation of this work has to do with the biases of \WP readership.
Given the popularity of the world's largest online encyclopedia, it is tempting to try to generalize our findings to large populations.
Yet, we must recall that this may not be true due to uneven access to the Internet and varying levels of digital literacy.
It could be the case, for example, that the groups with the largest increase in basic needs lack the capability to access \WP, and that we therefore lose the signals associated with this important population.

\xhdr{Future work}
A further interesting causal question asks in what activities people engage when forced to spend time at home.
Answering this would require drawing conclusions from the particular lockdown that we are observing in the context of \covid{} to general situations of restricted mobility.
One idea would be to treat \covid{} as an instrumental variable \cite{angrist1995identification}, \ie, a haphazard event that systematically nudges people to stay at home, while affecting people's interests only via the lockdown measures.
For instance, it would be interesting to explore whether \covid{} can be used as an instrumental variable to estimate what books people read, or what dishes they cook, when they are forced to, or---let the glass be half-full!---when they are given the chance to, spend more time at home.

\section*{Acknowledgments}

This work was partly supported by Swiss National Science Foundation grant 200021\_185043 and gifts from Google, Facebook, and Microsoft.
The project was kickstarted by a remote hackathon in the EPFL Data Science Lab, and we thank all participants for their important help: Akhil Arora, Alberto García Durán, Germain Zouein, Lars Klein, Liangwei Chen, Tiziano Piccardi, and Valentin Hartmann.

{\small
\bibliography{references}
}
\appendix

\section{Obtaining mobility changepoints}
\label{sec:appendix}



\xhdr{Mobility data} 
To estimate the effective lockdown dates, we use the mobility reports made available by Google and Apple. Google released community-level reports \cite{aktay2020google} indicating the daily percentage change in visits to predefined categories of places: \emph{Retail and Recreation} aggregates places like restaurants, caf\'es, shopping centers, \emph{Grocery and Pharmacy}, \emph{Parks}, \emph{Transit Stations} for public transport hubs, \emph{Workplaces}, and \emph{Residential} which estimates stay-at-home changes. The changes, for a given place on a given day, are reported in comparison to a baseline value, \ie, the median volume for the same day of the week computed across a five-week period between 3~January and 6~February 2020.
Similarly, Apple reports relative changes compared to a baseline volume measured on 13~January2020
along 3 dimensions of mobility: \emph{Driving}, \emph{Walking}, and \emph{Transit}. Combined, these represent nine types of time series capturing mobility behavior.

\xhdr{Handling the language-\vs-country mismatch}
Some languages (\eg, English, French, German) cannot easily be matched to one particular geographical area. For them, we collect the largest countries in terms of native speakers for which mobility data was available. We then produce an aggregate of the mobility data across countries weighted by the percentage of native speakers of the language in each country. 
For English, we aggregated: United States ($68.9\%$), United Kingdom ($16.1\%$), Canada ($5.8\%$), Australia ($5.4\%$), South Africa ($1.5\%$), Ireland ($1.2\%$), and New Zealand ($1.1\%$).
For French, we aggregated: France ($62.9\%$),
Canada ($10.1\%$), Cameroon ($9.2\%$), Belgium ($7.9\%$), Senegal ($4.3\%$), Benin ($3.8\%$), and Switzerland ($1.8\%$).
For German, we aggregated: Germany ($87\%$), Austria ($8.7\%$), and Switzerland ($6.3\%$).


\xhdr{Changepoint detection}
Changepoint detection
is the task of identifying state changes in 
time series. We can benefit from the large literature on offline changepoint detection to identify the effective time of lockdown based on changes in mobility time series. 
These techniques usually rely on two components: (1) a cost function assessing the quality of a particular signal segmentation and (2) a technique to search the space of possible segmentations, guided by the cost function
\cite{aminikhanghahi2017survey}.  
For more details, we refer to the review of \cite{truong2020selective}. In this work, we considered the cost functions and search algorithms available as part of the \texttt{ruptures} package.\footnote{\url{https://github.com/deepcharles/ruptures}}

Several different design choices can be made: (1)~different subsets of mobility time series can be selected and (2)~different changepoint algorithms can be employed.
We ran the pipeline with many different parameters and report the results in 
\Tabref{tab:changepoint}.
The standard deviations are due to changing parameters and are small, often below one day.
Indeed, for most countries, the changes in mobility are very clear, and different methods largely agree.


\begin{table}[t]
\small
\footnotesize
\centering
\begin{tabular}{l|cc||l|cc}
\toprule
Language & Mean & SD & Language & Mean & SD\\
\midrule
English     & 03/16 & 0.14 &   Dutch       & 03/16 & 0.97 \\
French      & 03/16 & 0.00 &   Norwegian   & 03/11 & 0.93 \\
German      & 03/16 & 0.07 &   Danish      & 03/11 & 0.06 \\
Korean      & 02/25 & 1.01 &   Swedish     & 03/11 & 1.02 \\
Japanese    & 03/31 & 1.07 &   Serbian     & 03/16 & 0.89 \\
 Finnish     & 03/16 & 0.14  &   Italian     & 03/11 & 0.00 \\
\bottomrule
\end{tabular}
\caption{Changepoint detection results averaged over different parameter choices. Standard deviations (SD) in days.}
\label{tab:changepoint}
\vspace{-5mm}
\end{table}

\end{document}